\documentclass[12pt,rmp,twocolumn,natbib,showkeys] {revtex4}

\usepackage[dvips]{graphicx}

\begin{document}

\title[Noise delayed switching of long Josephson junctions]
{Influence of length on the noise delayed switching of long Josephson junctions}

\author{\firstname{K.G.} \surname{Fedorov$^1$}}
\author{\firstname{A.L.} \surname{Pankratov$^1$}}
\author{\firstname{B.} \surname{Spagnolo$^2$}}
\affiliation{$^1$Institute for Physics of Microstructures of RAS,
GSP-105, Nizhny Novgorod, 603950, Russia} \affiliation{$^2$
Dipartimento di Fisica e Tecnologie Relative, Group of
Interdisciplinary Physics, Universit$\grave{a}$ di Palermo and
CNISM-INFM, Viale delle Scienze, I-90128, Palermo, Italy}

\begin{abstract}
The transient dynamics of long overlap Josephson junctions in the
frame of the sine-Gordon model with a white noise source is
investigated. The effect of noise delayed decay is observed for the
case of overdamped sine-Gordon equation. It is shown that this noise
induced effect, in the range of small noise intensities, vanishes
for junctions lengths greater than several Josephson penetration
length.
\end{abstract}
\date{\today}
 \maketitle
\newpage

\section{Introduction}

Josephson junctions are physical systems with nonlinear dynamics,
which are interesting to investigate both from experimental and
theoretical points of view. This is also in view of numerous
applications of superconductive devices based on Josephson
junctions, such as RSFQ devices, qubits, SQUIDs, etc.. [Barone $\&$
Patern$\grave{o}$, 1982; Likharev, 1986]. There are, in fact, a lot
of open problems in Josephson junction's dynamics, due to its
nonlinear character. For some devices, as the RSFQ, minimization of
the switching time is required for better performance [Pankratov
$\&$ Spangolo, 2004]. From this point of view, it is interesting to
investigate the influence of thermal noise on the statistical
properties of distributed Josephson junctions. It is also necessary
to note that, currently, all Josephson junctions are manufactured
with the use of optical and electron-beam lithography [Dorojevets,
2002; Makhlin \emph{et al.}, 2001], and can always be considered as
distributed. Moreover macroscopic quantum phenomena in long
Josephson junctions have attracted a lot of experimental and
theoretical work  recently [Weides \emph{et al.}, 2006; Mertens
\emph{et al.}, 2006; Alfimov $\&$ Popkov, 2006; Kim \emph{et al.},
2006; Fistul \emph{et al.}, 2003]. These junctions are characterized
by one or more dimensions longer than the Josephson penetration
length or depth [Barone $\&$ Patern$\grave{o}$, 1982].

It was shown in different physical systems that thermal fluctuations
can considerably increase the decay time of unstable states [Agudov
$\&$ Malakhov, 1995; Malakhov $\&$ Pankratov, 1996; Agudov $\&$
Malakhov, 1999] and the lifetime of metastable states [Mantegna $\&$
Spagnolo, 1996; Agudov $\&$ Spagnolo, 2001; Spagnolo \emph{et al.},
2004], producing a nonmonotonic behavior of these quantities as a
function of the noise intensity. These are the effects of noise
delayed decay (NDD) of unstable states and noise enhanced stability
(NES) of metastable states. Both noise induced effects are due to
the nonlinearity of the potential profile and are enhanced by the
inverse probability current. The NDD effect, in particular, consists
in the increase of the mean switching time (MST) of a Josephson
junction due to the influence of noise. The main interest to analyze
here long Josephson junctions (LJJ) is the existence of noise
delayed decay (NDD) effect for the restricted range of parameters,
which is important for practical applications.

\section{The model}

In the frame of the resistive McCumber-Stewart model [Barone $\&$
Patern$\grave{o}$, 1982; Likharev, 1986] the phase difference of the
order parameter $\varphi(x,t)$ of a long Josephson junction of the
overlap geometry (see Fig.~1) is described by the sine-Gordon
equation
\begin{eqnarray}
\beta\frac{\partial^2\varphi}{\partial t^2}+\frac{\partial\varphi}
{\partial t}-\frac{\partial^2\varphi}{\partial x^2}=i-\sin
(\varphi)+i_f(x,t) \label{PSGE}
\end{eqnarray}
\begin{figure}[t]
\begin{picture}(200,150)(10,10)
\centering\includegraphics[width=8cm,height=6cm]{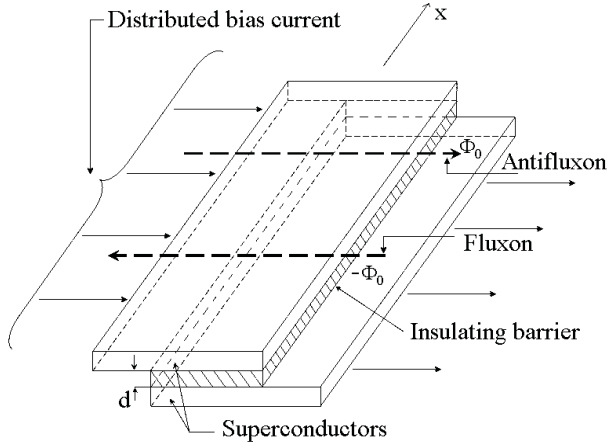}
\end{picture}
\parbox{220bp}
{\caption{The structure of distributed Josephson junction of the
"overlap" geometry.} \label{fig1}}
\end{figure}
with the following boundary conditions
\begin{equation}
\frac{\partial\varphi(0,t)}{\partial x}=
\frac{\partial\varphi(L,t)}{\partial x}=\Gamma. \label{Bndr}
\end{equation}
Here the time and the space are normalized to the inverse
characteristic frequency $\omega_{c}^{-1}$ and the Josephson
penetration length $\lambda_J$, respectively. The penetration length
gives a measure of the distance in which dc Josephson currents are
confined at the edges of the junction. $\beta=1/\alpha^2$ is the
McCumber-Stewart parameter, $\alpha = \omega_p/\omega_c$ is the
damping, with C the junction capacitance and $\omega_p = (2e
I_c/\hbar C)^{1/2}$ the plasma frequency corresponding to the
critical current $I_c$. The bias current density $i$ is normalized
to the critical current density of the junction, $i_{f}(x,t)$ is the
fluctuational current density, $\Gamma$ is the normalized magnetic
field, $L=l/\lambda_J$ is the dimensionless length of the junction.
Further, we will only consider the cases when $i > 1$ and $\Gamma =
0$, so in the potential profile there are no metastable states at
all. Also let us consider only the case of homogeneous bias current
distribution, when $i$ is constant along the junction. In the case
where the fluctuations are treated as white Gaussian noise with zero
mean, and the critical current density is fixed, its correlation
function is
\begin{equation}
\left<i_f(x,t)i_f(x',t')\right>=2\gamma \delta (x-x^{\prime})\delta
(t-t^{\prime}), \label{CorrFunc2}
\end{equation}
where
\begin{equation}
\gamma = I_{T} / (J_{c}\lambda_J) \label{gam2}
\end{equation}
is the dimensionless noise intensity, $J_{c}$ is the critical
current density of the junction, $I_{T}=2ekT/\hbar$ is the thermal
current, $e$ is the electron charge, $\hbar = h/2\pi$ with $h$ the
Planck constant, $k$ is the Boltzmann constant and $T$ is the
temperature. It is possible to obtain the expressions
(\ref{CorrFunc2}) and (\ref{gam2}) through the renormalization of
the general formula for the noise intensity, derived for the fixed
total critical current $I_{c}$ [Castellano \emph{et al.}, 1996;
Pankratov, 2002].

Initially, the whole phase "string" $\varphi(x,0)$ is located at the
position $\varphi_0$, so there are no vortices in the junction. The
dynamics of such a phase "string" is similar to that of a real long
string falling down on a tilted washboard potential addressing the
general problem of the diffusion of an elastic string on a tilted
periodic substrate [Cattuto \& Marchesoni, 1997]. For our case $i
> 1$, such initial phase is considered to be located in the
inflection point of the potential profile $\varphi_0= \pi / 2$. It
is possible to prepare such a state by a fast switching from the
starting superconductive state, corresponding to a metastable state,
to the resistive one, which becomes then the initial configuration
of the system. Due to the unstable character of the initial position
(see Fig.~2), the phase will begin the deterministic motion along
the potential profile. The MST is defined as the mean time of phase
$\varphi$ existence in the considered interval [$-\pi,\pi$]
[Malakhov $\&$ Pankratov, 2002], where $P(t)$ is the probability
that the phase is located in the initial interval
\begin{figure}[t]
\begin{picture}(200,150)(10,10)
\centering\includegraphics[width=8cm,height=6cm]{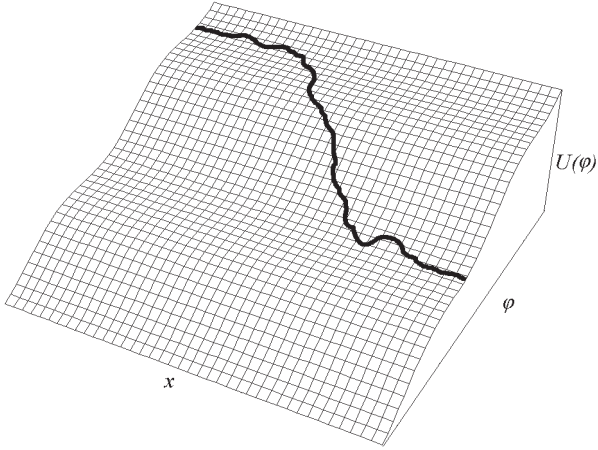}
\end{picture}
\parbox{220bp}
{\caption{Qualitative illustration of the potential profile
$u(\varphi)=1-\cos(\varphi)-i\varphi$, with $i = 1.2$, and phase
string $\varphi(x,t)$ in the process of escape from the initial
position.} \label{fig2}}
\end{figure}
\begin{eqnarray}\nonumber
\tau=\displaystyle{\int\limits_0^{+\infty} t w(t)dt} =
\int\limits_0^{+\infty} P(t)dt, \,
w(t)=-\displaystyle{\frac{\partial P(t)}{\partial t}}.
\end{eqnarray}
Here $w(t)$ is the probability density of the escape times. The
probability $P(t)$ is evaluated numerically in the following way: if
at the given time $t>0$ the realization of $\varphi(x,t)$ is within
the interval [$-\pi,\pi$], the probability for the corresponding
realization is one, otherwise it is zero. After $N$ realizations,
the average, both over the realizations and the spatial coordinate
$x$ from $0$ to $L$, is taken and, finally, the required probability
$P(t)$ is obtained.

Numerical solution of the sine-Gordon equation (\ref{PSGE}), with
boundary conditions (\ref{Bndr}), is performed by using the implicit
finite-difference scheme [Zhang, 1991], but taking care to insert
correctly the noise intensity [Federov $\&$ Pankratov, 2007].
Typical values of the simulation parameters are $\triangle x =
\triangle t = 0.1 - 0.02$, for the spatial and time discretization
steps, and $R = 10^3 - 10^4$ for the number of realizations.

\section{Noise delayed switching}

\begin{figure}[t]
\begin{picture}(200,150)(10,10)
\centering\includegraphics[width=8cm,height=6cm]{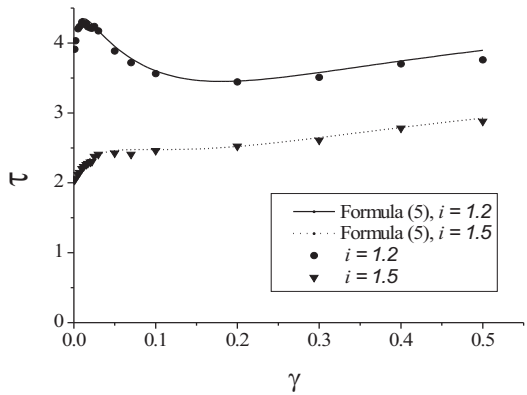}
\end{picture}
\parbox{220bp}
{\caption{MST versus the noise intensity $\gamma$. Comparison
between the theoretical results of Eq.~(\ref{MPF}) and the numerical
simulations of Eq.~(\ref{PSGE}), for two values of the bias current,
namely $i = 1.2, 1.5$. Here the dimensionless junction length is $L
= 0.1$.} \label{fig3}}
\end{figure}

\begin{figure}[h]
\begin{picture}(200,150)(10,10)
\centering\includegraphics[width=8cm,height=6cm]{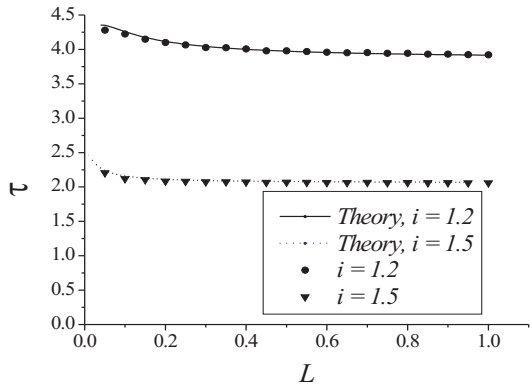}
\end{picture}
\parbox{220bp}
{\caption{MST versus the dimensionless junction length $L =
l/\lambda_J$. Comparison between the theoretical results of
Eq.~(\ref{MPF}) and the numerical simulations of Eq.~(\ref{PSGE}),
for the same values of the bias current $i$ of Fig.~3. Here the
dimensionless noise intensity is $\gamma = 0.005$. \vskip-0.1cm}
\label{fig4}}
\end{figure}

\begin{figure}[t]
\begin{picture}(200,150)(10,10)
\centering\includegraphics[width=8cm,height=6cm]{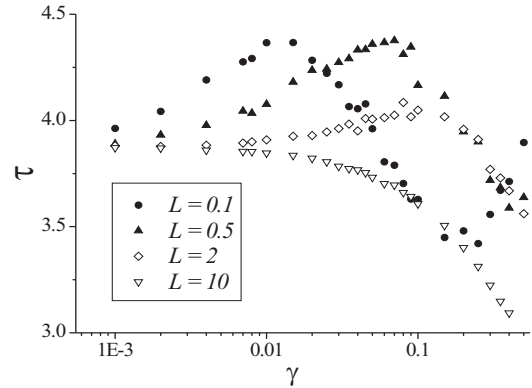}
\end{picture}
\parbox{220bp}
{\caption{Semilog plot of MST versus $\gamma$, for a long Josephson
junction with $i = 1.2$ and different junctions lengths, namely $L =
0.1, 0.5, 2, 10$.} \label{fig5}}
\end{figure}

\begin{figure}[h]
\begin{picture}(200,150)(10,10)
\centering\includegraphics[width=8cm,height=6cm]{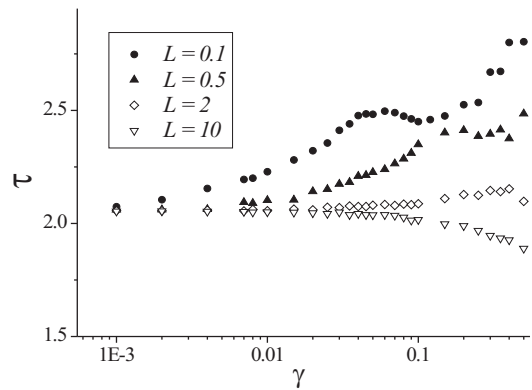}
\end{picture}
\parbox{220bp}
{\caption{Semilog plot of MST versus $\gamma$, for a long Josephson
junction with $i = 1.5$ and the same different junctions lengths of
Fig.~5.} \label{fig6}}
\end{figure}

First, it is interesting to check the limiting transition of a long
junction to a point one for small junction lengths $L\ll 1$. For a
point junction, in the case of large damping $\beta \ll 1$, the MST
was found analytically [Malakhov $\&$ Pankratov, 1996] for an
arbitrary value of the noise intensity. That formula, however, was
obtained for the case of constant critical current $I_c$, and the
noise intensity in that case was: $\gamma_s=I_T/I_c$, where
$I_c=J_c\lambda_J L$. This means that we have to scale the noise
intensity as $\gamma_s=\gamma/L$, with $\gamma$ given by
Eq.(\ref{gam2}). We get, therefore, the following closed expression
for the MST in a long Josephson junction with constant critical
current density
\begin{eqnarray} \tau =  \displaystyle{\frac{L}{\gamma}\left\{
\int_{\varphi_0}^{\varphi_2 }e^{-f(x)L/\gamma }
\int_{\varphi_1}^xe^{f(\varphi)L/\gamma }
d\varphi dx \right.}  \nonumber\\
\displaystyle{\left.+\int_{\varphi_1}^{\varphi_2}e^{f(\varphi)L/\gamma}
d\varphi\cdot \int_{\varphi_2 }^\infty e^{-f(\varphi)L/\gamma}
d\varphi \right\}}, \label{MPF}
\end{eqnarray}
with
\begin{equation}
f(\varphi) = \cos\varphi + i \varphi \; .
\end{equation}
In Eq.~(\ref{MPF})) $\varphi_0$ is the coordinate of the initial
delta-shaped distribution, and $\varphi_{1,2}$ are the boundaries of
the interval, delimiting the potential well or metastable state of
the superconductive state, before the fast switching. To prevent
misunderstanding with the use of such a renormalization procedure,
one should consider the general expression for the noise intensity
for fixed critical current [Castellano \emph{et al.}, 1996;
Pankratov, 2002] and substitute the required bias current density in
it. In Figs.~\ref{fig3},~\ref{fig4} the behavior of MST as a
function of the dimensionless noise intensity $\gamma$ and the
junction length $L$, respectively,  is shown for two values of the
bias current. The agreement between the theoretical results,
obtained from Eq.~(\ref{MPF}), and the numerical simulations of
Eq.~(\ref{PSGE}), for a long junction, is very good not only in the
limiting case $L\ll 1$, but even up to $L\sim 1$. In
Figs.~\ref{fig5},~\ref{fig6} the semilog plot of the MST versus the
noise intensity $\gamma$,
for different junction lengths, is shown. The NDD effect is present
for junction lengths up to $L \gtrsim 5$, while for greater lengths
it completely vanishes. This peculiarity is very important for RSFQ
devices. This disappearance of NDD is due to the effective decrease
of the noise intensity in comparison with the point junction.

For the same temperature level, the effective noise intensity
$\gamma_s=\gamma/L$ for the long Josephson junction will be smaller
by a factor $1/L$ in comparison with the point junction. So, for LJJ
with rather large lengths ($L>5$) the noise intensity will get out
of the area of the NDD effect.

For example, from our simulations, we find that for great values of
the noise intensity $\gamma \sim 100$, the NDD will exists even for
LJJ with lengths $L \sim 10$, but such a range of $\gamma$ is not
interesting from practical point of view. In other words, for large
lengths of the distributed Josephson junctions, the random force
$F_T \sim \sqrt{\gamma}$ becomes negligibly small in comparison with
the deterministic force, caused by the potential profile
$u(\varphi)=1-\cos(\varphi)-i\varphi$.

The standard deviation (SD) $\sigma$ of the switching time is
defined as
\begin{eqnarray}\nonumber
\sigma = \sqrt{ \langle t^{2} \rangle - \tau^{2}}, \langle t^2
\rangle = \int^\infty_0 t^2 w(t) dt.
\end{eqnarray}
For small noise intensities we can obtain an expression of
$\sigma(\varphi_0)$ for a long Josephson junction, by using the
noise intensity renormalization, as before in Eq.~(\ref{MPF}), and
the asymptotic expression derived for a point Josephson junction
[Pankratov $\&$ Spagnolo, 2004]
\begin{eqnarray}\label{sg}
\sigma(\varphi_0)=\frac{1}{\omega_c}\sqrt{(2\gamma / L)
\left[F(\varphi_0)+f_3(\varphi_0)\right]+ .. },\\
\begin{array}{ccc}
F(\varphi_0)&=&f_1(\varphi_2)f_2(\varphi_2)-2f_1(\varphi_2)f_2(\varphi_0)
\nonumber \\
& &+
f_1(\varphi_0)f_2(\varphi_0)+\frac{f_1(\varphi_2)-f_1(\varphi_0)}{(i-\sin(\varphi_0))^2},
\nonumber \\
f_1(x)&=&\frac 2{\sqrt{i^2-1}}
\arctan\left( \frac{i\tan(x/2)-1}{\sqrt{i^2-1}}\right), \\
\nonumber
f_2(x)&=&1/({2(i-\sin{x})^2}), \\
f_3(\varphi_0)&=& \int_{\varphi_0}^{\varphi_2}
\left[\frac{\cos(x)f_1(x)}{(\sin(x)-i)^3}-\frac{3}{2(\sin(x)-i)^3}
\right]dx.
\end{array}
\end{eqnarray}
In the following Fig.~7 the SD of the switching time versus the
junction length L is shown, for two values of the bias current,
namely $i = 1.2, 1.5$.
\begin{figure}[h]
\begin{picture}(200,150)(10,10)
\centering\includegraphics[width=8cm,height=6cm]{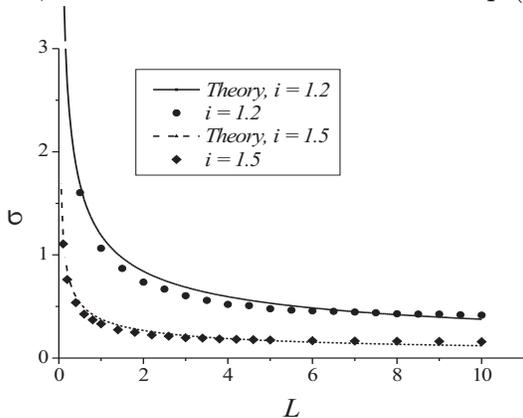}
\end{picture}
\parbox{220bp}
{\caption{SD versus the dimensionless junction length L for two
values of the bias current, namely $i = 1.2, 1.5$. The theoretical
results of Eq.~(\ref{sg}) are compared with the numerical
simulations of Eq.~(\ref{PSGE}). Here the noise intensity is $\gamma
= 0.01$.}
\label{fig7}}
\end{figure}
The theoretical results of Eq.~(\ref{sg}) show a quite good
agreement with the numerical simulations of Eq.~(\ref{PSGE}).
\begin{figure}[h]
\begin{picture}(200,150)(10,10)
\centering\includegraphics[width=8cm,height=6cm]{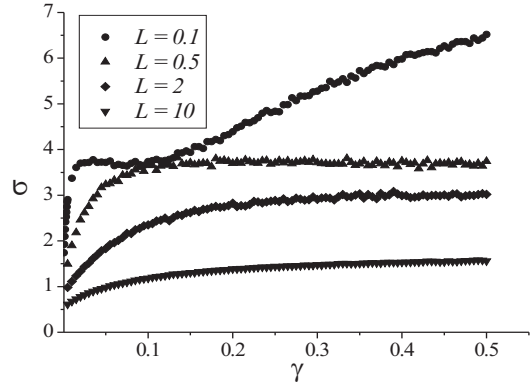}
\end{picture}
\parbox{220bp}
{\caption{Standard deviation versus the dimensionless noise
intensity $\gamma$ for four junction length, namely $L =
0.1,0.5,2,10$. Here the bias current is $i = 1.2$.}
\label{fig8}}
\end{figure}
\begin{figure}[h]
\begin{picture}(200,150)(10,10)
\centering\includegraphics[width=8cm,height=6cm]{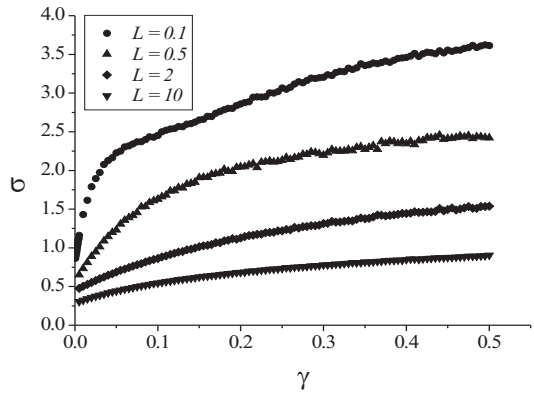}
\end{picture}
\parbox{220bp}
{\caption{Standard deviation versus the dimensionless noise
intensity $\gamma$ for the same junction lengths of Fig.~8. Here the
bias current is $i = 1.5$.
\vskip-0.5cm} \label{fig9}}
\end{figure}
In Figs.~8,~9 the numerical simulations of the SD versus the noise
intensity from Eq.~(\ref{PSGE}), for two values of the bias current
and for different junction lengths, are shown. All the behaviors of
Figs.~7,~8,~9 are evaluated in a parameter range which is
characteristic for RSFQ devices. The SD decreases with the length of
the junction. This is because the dynamics controlling the switching
event goes from a noise-induced regime, at very short junction
lengths, to a deterministic regime, caused by the decrease of the
effective noise intensity for very long junctions. The noisy regime
is well visible in Figs.~8,~9 for the case $L = 0.1$, for which the
$\sigma$ increases with the noise intensity. The deterministic
regime, dominated by the potential profile, produces the asymptotic
constant values of the SD (the curves for $L = 0.5, 2, 10$ in
Figs.~8,~9). We note that for a higher bias current (Fig.
\label{fig9}) the SD reaches lower asymptotic values, as we expect.

\section{Conclusions}

In this work a study of the transient dynamics of long Josephson
junctions, with bias currents greater than the critical current, is
presented. We have found that, for the case of constant critical
current density, the mean switching time has a maximum for small
noise intensities and junction lengths. This is the noise delayed
decay effect. This effect disappears for junctions with
dimensionless lengths $L \gtrsim 5$, which can be important for the
design of RSFQ devices based on Josephson junctions. The
disappearance of NDD effect is explained by the decrease of the
effective noise intensity with the length $L$ of the junction, by
considering the scaling factor $1/L$. As a consequence the system
gets out from the range of $\gamma$ values suitable to observe the
noise delayed decay effect. Finally we observe for the MST and its
SD, in the range of small noise intensities, a good agreement
between the theoretical behaviors (Eqs.~(\ref{MPF}) and~(\ref{sg}))
and those obtained from numerical simulations of Eq.~(\ref{PSGE}).

\end{document}